\documentclass[10pt,twocolumn,prd]{revtex4}

\usepackage{amssymb}

\usepackage{graphicx}

\begin{document}
\title{Ranking nodes according to their path-complexity}


\author{Francesco Caravelli}

\affiliation{Department of Computer Science,\\ UCL, Gower Street, London, WC1E6BT, UK\\ \textit{and}\\
Invenia Technical Computing, 135 Innovation Dr., Winnipeg, MB R3T 6A8, Canada
}

\begin{abstract}
Thermalization is one of the most important phenomena in statistical physics. Often, the transition probabilities between different states in the phase space is or can be approximated by constants. In this case, the system can be described by Markovian transition kernels, and when the phase space is discrete, by Markov chains. In this paper, we introduce a macroscopic entropy on the states of paths of length $k$ and, studying the recursion relation, obtain a fixed point entropy. This analysis leads to a centrality approach to Markov chains entropy.
\end{abstract}

\maketitle

\section{Introduction}

Physical systems require very often different descriptions at the micro and macro scale. It is the case for instance in systems which exhibit emergent phenomena, and for systems which undergo a phase transition. In this case, one could argue that the degrees of freedoms change with the scale effectively, and thus phase space counting should be different depending on the lens with which one look at the system. This line of thinking has been very fruitful in the last century, since the very initial work of Gell-Man and Low on the renormalization group. The concept of emergence, in particular, has enlighted many physical phenomena, giving them in the first place both a renewed appeal from the new interpretation. With this same line of reasoning, dynamical systems can often exhibit correlations which are not only time dependent, but that at short time scales with respect to thermalization typical time scale, exhibit different behaviors.


The introduction of a macroscopic entropy functional for statistical systems has been introduced by Lloyd and Pagels in \cite{Lloyd},
and at the same time by Lindgren \cite{Lindgren}. Lloyd and Pagels showed that the depth of a Hamiltonian system is proportional to the difference between the system and the coarse grained entropy. This paper introduced the concept of ``thermodynamic depth''. If $p_i$ is the probability that a certain system arrived at a macroscopic state $i$, then the thermodynamic depth of that state is proportional to $\ln(p_i)$. This implies that the average depth of a system, the complexity, is proportionall to the Shannon entropy, or the Boltzman entropy. In addition, it has been shown in \cite{Lloyd} that the only functional that is continuous, monotonically increasing with system size, and is extensive is the Boltzman functional up to a constant \cite{vinfo}. One can show that such argument is true also for \textit{macroscopic states}, described by trajectories $i_1\rightarrow i_2 \rightarrow i_3 \cdots \rightarrow i_n$. In this case, the thermodynamic depth of this state is given by $- \alpha \log p(i_1,i_2,i_3, \cdots | i_n)$. In general, the average depth of a system with many macroscopic states can be very large. In fact, it has been shown in \cite{Lindgren} that the macroscopic entropy defined by:
\begin{equation}
S_m=\sum_{i_1 i_2 \cdots i_m} -p_m(i_1 i_2 \cdots i_m) \log(p_m(i_1 i_2 \cdots i_m))
\end{equation}
is monotonically increasing, i.e. $\Delta S_m=S_m-S_{m-1}\geq 0$, and $\Delta S_m-\delta S_{m-1}\leq 0$. It has been also shown that, if in general the macrostate is described by a string of length $L$, one can obtain a finite specific thermodynamic depth,
$$\mu(\chi) =\lim_{m\rightarrow \infty} \frac{S_{m}}{m}, $$
with $\chi$ being the infinite string.
The idea of thermodynamic depth has inspired Ekroot and Cover to introduce the entropy of Markov trajectories in \cite{Ekroot}.
If $P_i$ denote the $i$th row of a Markov transition matrix, one can define the entropy of a state $i$ as:
\begin{equation}
H(P_i)=-\sum_j M_{ij} \log(M_{ij})
\end{equation}
with $M_{ij}$ being the Markov operator. If one introduces the probability of a trajectory going from $i$ to $j$ as $p_{ij}$, then, the macroscopic entropy of the Markov trajectory is given by:
\begin{equation}
H_{ij}=-\sum_j P(t_{ij}) \log(P(t_{ij})).
\end{equation}
For Markov chains, one has that $p_{ij}=\sum_{k_1,k_2,\cdots,k_n} M_{i k_1}\cdots M_{k_n j}$, which thus leads to a recurrence relation:
\begin{equation}
H_{ij}=H(P_{i})+\sum_{k\neq j} P_{ij} H_{kj},
\end{equation}
which follows from the chain rule of the entropy, and allows to calculate a closed formula for $H_{ij}$ in terms of the entropy of the nodes, that we will call $\ ^1 S_{i}$, and the asymptotic, stationary distribution of the Markov chain, $\pi$. 

Over the last decade, a huge effort has been devoted to understanding processes on networks \cite{dyncn}, understanding their statistical properties, as interactions very often occur on nontrivial network topologies, as for instance scale free or small world networks, called complex networks.
With this widespread interest in networks, the study of global properties of graphs and graph ensembles has given a renewed impetuous to the study of entropies on graphs.  
In general, in analogy with what happens for Markov chains, one is interested in quantifying their complexity by means of information theory approach. Since for strongly connected graphs, the transition kernel, given by $M=D^{-1} A$, with $A$ being the adjacency matrix of the graph and $D$ being the diagonal matrix of degree with $D_{ii}=\sum_{j} A_{ij}$. If $M$ is an ergodic operator (which depends on the topological properties of the underlying graph), one can study operators based on the asymptotic properties of a random walk.

The dynamics and the structure of many physical networks, such as those involved in biological, physical, economical and technological systems, is often characterized by the topology of the network itself. 

  In order to quantify the complexity of a network, several measures of complexity of a network have been introduced, as for instance in \cite{compnet1}, studying the entropy associate to a certain partitioning of a network.
The standard Boltzmann entropy per node was defined as the transition kernel of a random walk in 
\cite{compnet2}. In general, in complex networks, one is interested in the average complexity of an ensemble of networks of the same type, as for instance Erd\`{o}s-Renyi or Watts-Strogats and Barab\'{a}si-Albert random graphs. Along these lines in particular, we mention the entropy based on the transition kernel of Anand and Bianconi \cite{Anand}. One can in fact write the partition function of a network ensemble subject to a micro-canonical constraint (the energy) and then, given the probability of certain microcanonical ensemble, calculate its entropy, similarly to what proposed in \cite{entrg} for random graphs.

In general, an entropy of a complex network can be associated from a test particle performing a diffusion process on the network, as in \cite{entratediff}; for scale free networks, it is found that the entropy production rate depends on the tail of the distribution of nodes, and thus on the exponent of the tail.

Along these lines, in \cite{Laplacian} a von Neumann entropy based on the graph laplacian has been introduced, merging results inspired from pure states in Quantum Mechanics, and networks, and finding that the von Neumann entropy is related to the spectrum of the laplacian \cite{Anand}. In particular, it has been shown that many graph properties can be identified using this laplacian approach. 

A huge body of work has been done by the group of Burioni and Cassi, in defining the statistical properties of graphs for long walks, for instance using a Heat Kernel approach to study spectral and fractal dimensions of a graph, finite and infinite (see \cite{BurioniCassi} and references therein).

In general, these approaches rely on a local operator (transition kernels, laplacians) with support on the graph. Therefore, if one is interested in knowing macroscopic properties of the graph, is indeed forced to use non-local operators. In addition to the theoretical interest of describing the macroscopic properties of a graph in terms of information theory quantities, it is important to remark that very often these have important applications in classifying systems according to their topological properties. For instance, in \cite{EntNSR} it has been shown that graph entropy can be used to differentiate and identify cancerogenic cells.
In particular, \cite{EntNSR} shows the importance of studying  entropies based on the non-local (macroscopic) properties of a network, as for instance the \textit{higher-order} network entropy given by

\begin{equation}
S^{(n)} = -\sum_j K_{ij}^{(n)} \log (K_{ij}^{(n)}),
\end{equation}
with $K_{ij}^{(n)}$ satisfying an approximate diffusion equation at $n$th order, 
$$K_{ij}^{(n)}\approx e^{M}+O(M^n). $$

In addition to the approaches just described, one could think of using, instead of the diffusion kernel above, a node-entropy based on diffusion as $S_i= \sum_j M_{ij}^k log(M^k_{ij})$. It is easy to see, however that for $k\rightarrow \infty$, if the operator is ergodic, the asymptotic entropy is independent from the initial state: it easy a known fact that if $M$ has a unique Perron root, $(M^k)_{ij}\approx \pi(j)+(N^k)_{ij}$, where $N$ 
is a Nilpotent operator such that $\lim_{k\rightarrow \infty} N^k=0$.
The same happens for the diffusion kernel at long times: in this case the diffusion kernel approaches the asymptotic distribution, which indeed has forgotten from which node the diffusion started. With the aim of retaining the information on the node, we introduce the entropy on the paths originating at a node which, as we shall show, has very interesting asymptotic properties for long walks. In the next section we describe the construction of the 
non-local entropy, and an application to random graphs and fractals. Conclusions follow.

\section{A node entropy based on paths}
We shall start by introducing some basic definitions. Let us consider a Markov operator $M_{ij}$ on $N$ states, i.e.$1\le\ i,j\ \le N$ ,
such that:
\begin{equation}
\sum_{j=1}^N M_{ij} =1.
\end{equation}
with $0\le M_{ij}\le 1$. Entropy gives a measure of how mixing are some states, i.e. how much one state is related to the other states $j$. We can the define the following quantity:

\begin{equation}
\ ^1S_i=-\frac{1}{N}\sum_{j=1}^N M_{ij} \log(M_{ij})
\label{entropy}
\end{equation}
that for reasons it will be clear soon, we call first order entropy\footnote{We assume that $0\cdot log(0)=0$.}.

It is clear, however, the this definition is purely local, i.e., the mixing defined by the first order entropy is a local concept, as the it gives a sense of how much mixing there is at the second step in a Markov process for each node.
Generalizing this entropy for longer times, i.e. when the $M$ operator is applied several times, is not obvious.  One obstruction might be given, in fact, by the ergodicity of the operator:
\begin{equation}
\lim_{n\rightarrow \infty} M^n=M^*;
\end{equation}
in this case, the operator $M^*$ is trivial, in the sense that each row of $M^*$ is identical, due to the ergodic theorem, and thus eqn. (\ref{entropy}) is non-trivially generalizable if one wants to assign a ranking to each node.
In the approximation of long walks, if one used the operator $M^*$, eqn. (\ref{entropy}) would become:
\begin{eqnarray}
\ ^*S_i &=&\lim_{k\rightarrow \infty}- \frac{1}{N} \sum_j (\sum_{j_1,\cdots ,j_k=1}^N M_{i j_1}\cdots M_{j_{k} j} )\cdot \nonumber \\
&\cdot& \log(\sum_{j_1,\cdots ,j_k=1}^N M_{i j_1}\cdots M_{j_{k} j})\nonumber \\
&=&-\frac{1}{N}\sum_{j=1}^N M^*_{ij} \log(M^*_{ij})=\ ^* S_i \equiv \ ^*S .
\label{entropytr}
\end{eqnarray}
As a result of the ergodic theorem, the entropy evaluated on asymptotic  states  is independent from the initial condition.

However, here we argue that there is a definition of entropy which indeed depends on the initial condition, which is the macroscopic entropy evaluated on the space of trajectories. We thus introduced the following entropy on the paths a Markov particle went through after $k$-steps, or $k$th order entropy:
\begin{equation}
\ ^kS_i=-P(k) \sum_{i_1,\cdots ,i_k=1}^N M_{i i_1} \cdot \cdot M_{i_{k-1} i_k}  \log(M_{i i_1} \cdots M_{i_{k-1} i_k}),
\label{entropyk}
\end{equation}
where $P(k)$ is a factor which depends only on $k$, and is used to keep the entropy finite in the limit $k\rightarrow \infty$. We will first assume that $P(k)$ does not depend on any other parameter; this choice easily leads to the factor $P(k)=\frac{1}{N k}$. Following the discussion in \cite{Lindgren}, it is easy to show, after having defined the $\delta$ operation on the entropy on paths of length $k$, 
\begin{equation}
\delta \ ^k S_i \equiv \ ^k S_i-\ ^k S_i,
\label{delta}
\end{equation}
that $\forall i$ and $\forall k$, $\delta \ ^k S_i \ge 0$ and $\delta^2 \ ^k S_i \le 0$.\ \\\ \\
This  implies also that this definition of macroscopic entropy has good asymptotic properties, i.e. that $\forall i$, $\exists$ a unique $\ ^*S_i$ such that $$\lim_{k\rightarrow \infty} \ ^k S_i = \ ^* S_i .$$  

In order to better interpret this non-local entropy, we introduce the following notation. We denote with $\{\gamma\}_k$, a path, a string of states of length k, $\{ i_1 \cdots i_k\}$ and with $\{\ _i\gamma_j\}$ an infinite string of states of the form $\{i \cdots j\}$. We then denote as $M(\{\{\ _i\gamma_j\}\}_k)$ the ordered product $$\prod_{\{\ _i \gamma_j \}_k}=M_{i i_1} M_{i_1 i_2} \cdots M_{i_{k-2} j},$$ and with 
$M(\{\ _i\gamma_j\})$ the infinite product, 
$$\prod_{\{\ _i \gamma_j \}}=M_{i i_1} M_{i_1 i_2} \cdots M_{i_{\infty} j}.$$
We also denote with $\sum_{\{\{\ _i\gamma_j\}_k\}}$ ( $\sum_{\{\{\ _i\gamma_j\}\}}$ ), the sums over all possible paths of length $k$ (infinite) starting in $i$ and ending in $j$, and $\sum_{\{\ _i\gamma\}_k}$ the sum over all possible paths of length $k$ (infinite) starting at $i$. 
We can then write, compactly (setting temporarily $P(k)=1$):

\begin{equation}
\ ^k S_i= -\sum_{\{\{\ _i\gamma\}_k\}} M(\{\ _i\gamma\}_k) \log(M(\{\ _i\gamma\}_k)).
\end{equation}
It is now easy to see that this can be written in terms of products:
\begin{equation}
e^{-\ ^k S_i}= \prod_{\{\{\ _i\gamma\}_k\}} M(\{\ _i\gamma\}_k)^{M(\{\ _i\gamma\}_k)},
\end{equation}
which gives an idea of how fast this product can grow as a function of $k$. In order to set the stage for what follows,
let us consider the simpler case of a 2 dimensional Markov chain with transition probabilities parametrized by two positive parameters, $0 \le a,b \le 1$:
\[ M= \left( \begin{array}{cc}
a & 1-a\\
1-b & b \end{array} \right),\]
and our aim is now to use a recursion relations for the infinite trees in order to calculate the exact values for $\ ^*S_1$ and $\ ^*S_2$.  This can be easily generalized, and in fact there is no obstruction to calculate this for generic Markov matrices;  as we will see shortly the result is independent on the dimensionalilty of the matrix.

Let us thus consider the entropies for $\ ^{k+1} S_1$ and $\ ^{k+1} S_2$. These can be written recursively, for $k$, as 
\begin{eqnarray}
\ ^{k+1}S_1&=&\frac{a}{k+1} \sum_{\{\{\ _1 \gamma\}_k\}} M(\{\ _1 \gamma\}_k) \log(a M(\{\ _1 \gamma\}_k)) \nonumber\\
&+&\frac{(1-a)}{k+1} \sum_{\{\{\ _2 \gamma\}_k\}} M(\{\ _2 \gamma\}_k) \cdot \nonumber \\
&\cdot& \log((1-a) M(\{\ _2 \gamma\}_k)) \nonumber\\
\ ^{k+1}S_2&=&\frac{(1-b)}{k+1} \sum_{\{\{\ _1 \gamma\}_k\}} M(\{\ _1 \gamma\}_k) \cdot \nonumber \\
&\cdot& \log( (1-b) M(\{\ _1 \gamma\}_k)) \nonumber\\
&+& \frac{b}{k+1}  \sum_{\{\{\ _2 \gamma\}_k\}} M(\{\ _2 \gamma\}_k) \log(b M(\{\ _2 \gamma\}_k))\nonumber \\
\end{eqnarray}
Now we can use the properties of logarithms, and the fact that 
$$\sum_{\{\{\ _r \gamma\}_k\}} M(\{\ _r \gamma\}_k) =1\ \forall r,$$ following from the fact that the matrix is stochastic.

We can at this point separate the various terms, obtaining:
\begin{widetext}
\begin{eqnarray}
& \ ^{k+1}S_1= \frac{1}{k+1}  a\log(a) + a \frac{k}{k+1}  \ ^{k}S_1 + \frac{1}{k+1} (1-a) \log(1-a) + (1-a)\frac{k}{k+1} \ ^{k}S_2 ,\nonumber \\
& \ ^{k+1}S_2=\frac{1}{k+1} b \log(b) + b\frac{k}{k+1}  \ ^{k}S_1 + \frac{1}{k+1} (1-b) \log(1-b) + (1-b)\frac{k}{k+1} \ ^{k}S_2 ,\nonumber \\
\end{eqnarray}
\end{widetext}
and thus we reach the following recursive equation:
\begin{equation}
\ ^{k+1} \vec S = \frac{1}{k+1}( k\ M\ \ ^{k}\vec S + \ ^1\vec S )
\end{equation}
where we see that the first order entropy enters:
$$\ ^1\vec S= \left( \begin{array}{c}
a \log(a)+ (1-a) \log(1-a)  \\
b \log(b)+ (1-b) \log(1-b)  \end{array} \right)\equiv \left( \begin{array}{c}
\ ^1 S_1 \\
\ ^1 S_2   \end{array} \right)$$
In doing the calculation, we observe that now we have a generic formula, which depends only on the Markov operator $M$. Due to the linearity of the recursion relation, it is easy to observe that it is independent from the dimensionality.
We observe that the $k=1$ case can be taken into account by defining $\ ^0\vec S=\vec 0$. 

For generic $k$, this equation can be written as:
\begin{equation}
\ ^{k+1}\vec S=\frac{1}{k+1}(\sum_{n=0}^k M^n) \ ^1 \vec S.
\label{recursion}
\end{equation}
First of all we notice that the limit 
\begin{equation}
\tilde M=\lim_{k\rightarrow \infty} \frac{1}{k+1} \sum_{n=0}^k M^n 
\end{equation}
is well defined, and so is its Ces\'{a}ro mean. To see this, we see that $M$ is a positive bounded operator, $||M||\le 1$. Thus, 
\begin{equation}
|| \tilde M ||\le \lim_{k\rightarrow \infty} \frac{1}{k+1} \sum_{n=0}^k || M^n || \le \lim_{k\rightarrow \infty} \frac{1}{k+1} \sum_{n=0}^k 1 = 1
\end{equation}
By the Ces\'{a}ro mean rule, we have then that 
\begin{equation}
\ ^*\vec S=\lim_{k\rightarrow \infty} \frac{1}{k+1}(\sum_{n=0}^k M^n) \ ^1 \vec S= \lim_{k\rightarrow \infty}  M^k \ ^1\vec S=M^* \ ^1 \vec S
\end{equation}
and thus we discover that also for this entropy, $\ ^* S_i \equiv \ ^* S$, with $\ ^* S=\sum_j M^*_{ij}\ ^1 S_j$, and thus we failed yet to distinguish the entropy of the paths for each single node. It is easy to realize that this is due to the normalization factor, $1/k$, which thanks to the Ces\'{a}ro rule leads to a different result we were looking for to begin with.

In order to do improve the counting, then, we can assume that now the normalization factor $P(k)$ depends on an extra parameter $\epsilon$, $P(k,\epsilon)$. In particular, 
we will be interested in the natural choice of contractions, i.e. $P(k,\epsilon)=\epsilon^k$, as this choice has nice asymptotic behavior and has a straightforward interpretation in terms of path lengths, as we will see after. We thus consider the following entropy functional:

\begin{equation}
\ ^k_\epsilon S_i=-\frac{\epsilon^{k-1}}{N}\sum_{i_1,\cdots ,i_k=1}^N M_{i i_1} \cdots M_{i_{k-1} i_k}  \log(M_{i i_1} \cdots M_{i_{k-1} i_k}).
\label{entropyk}
\end{equation}
This formula is now defined in terms of an extra parameter $\epsilon$; we will now show that thanks to the recursion rule, one can find a closed formula at $k\rightarrow \infty$. 

\subsection{Fixed point for $\epsilon$-Path entropy and centrality}

Having gained experience on how to write the recursion rule in the previous section, we promptly modify the recursion rule in order to account for the normalization $P(k)=\epsilon^k$.
Thus, following the same decomposition in order to find the recurrence rule, we find:
\begin{widetext}
\begin{eqnarray}
& \ ^{k+1}_\epsilon S_1= \epsilon^{k}  a\log(a) + \epsilon\ a   \ ^{k}_\epsilon S_1 + \epsilon^{k} (1-a) \log(1-a) + \epsilon\ (1-a) \ ^{k}_\epsilon S_2 ,\\
& \ ^{k+1}_\epsilon S_2=\epsilon^{k} b \log(b) + \epsilon\ b   \ ^{k}S_1 + \epsilon^{k} (1-b) \log(1-b) +  \epsilon\ (1-b) \ ^{k}_\epsilon S_2 ,
\end{eqnarray}
\end{widetext}
which leads to the following closed formula for the recursion, in terms of $M$, $\ ^1 \vec S$ and $\epsilon$:
\begin{equation}
\ ^{k+1} \vec S_\epsilon = \epsilon^{k} ( \ \frac{M}{\epsilon^{k-1}}\ \ ^{k}\vec S + \ ^1\vec S ).
\end{equation}
Writing down all the terms, recursively, we find:
\begin{equation}
\ ^{k+1} \vec S_\epsilon = \sum_{n=0}^{k} \epsilon^n M^n  \ ^1\vec S,
\end{equation}
and, realizing that we can now take the limit $k \rightarrow \infty$  safely,:
\begin{equation}
\ ^{*} \vec S_\epsilon = \frac{ 1}{I- \epsilon M} \ ^1\vec S ,
\label{entfin}
\end{equation}
which is finite if $\epsilon<1$, and is the main result of this paper. A compact way of rewriting equation in eqn. (\ref{entfin}), is by multiplying and dividing by $\epsilon$, and writing the entropy in terms of the matrix resolvent $R(\frac{1}{\epsilon},A)$.
\begin{equation}
\ ^{*} \vec S_\epsilon = \frac{1}{\epsilon}\frac{ 1}{\frac{1}{\epsilon}I-  M} \ ^1\vec S =\frac{1}{\epsilon}R(\frac{1}{\epsilon},M) \ ^1\vec S
\label{resolvent}
\end{equation}

We thus now see that we have traded the infinity for a ``forgetting'' parameter $\epsilon$, which adds a further variable to the analysis, and which might seem puzzling at first. In particular, we do recognize that this operator has been widely used in several fields, which is reassuring. In fact, the entropy just introduced resembles several centrality measures on network, as for instance the Katz centrality, although applied to a vector which is different than a column of ones, but has the entropy calculated at the first order for in each node. In particular, we realize that the resolvent is often used for measuring correlations (for instance in \cite{Griffith}).

It is worth make few comments regarding eqn. (\ref{entfin}). It is striking, as often it happens, that it is indeed easier to calculate this path entropy, thanks to the fixed point, for an infinite number of steps, rather than a finite number. We can in fact easily convince ourselves that calculating all the paths of length $k$ on a graph can grow as $n^k$ in the worst case, which can be a rather big number for fairly small graphs after few steps. Using the formula above, one can calculate this entropy by merely inverting a matrix which, apart from being the fortune of several search engines and big data analysts, and although being slow in some cases or can have convergence problems, can be done also for large matrices (and thus graphs), which is rather convenient. If this was not enough, one can however also tune the parameter $\epsilon(\langle l \rangle)$ in order calculate the entropy (on average) after a finite number of steps $\langle l \rangle$, as we shall soon show. 

In general, one could generalize this formula by refining on the type of paths one is interested of summing on (for instance, self-avoiding loops, closed random walks). Although this approach is definitely feasible, it is hard, at the end of the computation, to find a closed formula at the fixed point. The reason is that, by summing on all possible indices, the equations can be written in terms of matrix multiplication of the Markov transition kernel, thus simplifying the final equation.  

As a final remark, we note that, differently from the approach of \cite{Ekroot}, we define the macroscopic entropy not on Markov trajectories defined by a source node $i$ and a destination node $j$, but indeed are aimed at studying the path complexity attached to a node $i$, given by all the possible paths which can be originated from it.

\subsection{Interpreting the $\epsilon$-parameter}
The introduction of a renormalization parameter $\epsilon$, able to keep the entropy finite in the asymptotic limit, and at the same time pertaining the information on the originating node, might seem puzzling at first.
In general, as we shall show now, one can associate the parameter $\epsilon$ to the average length path to be considered. In fact, one can write the average path length, recursively, and obtain the formula:

\begin{eqnarray}
\langle l \rangle &=& \sum_{i=0}^{\infty} l \epsilon^l=\epsilon \partial_\epsilon  \sum_{i=0}^{\infty} \epsilon^{l}\nonumber \\
&=&  \frac{\epsilon}{(1-\epsilon)^2}
\end{eqnarray}
and under the assumption that $\epsilon(0)=0$, one can obtain the roots of this equation for $\epsilon$ as a function of $\langle l \rangle$, 
\begin{equation}
\epsilon(\langle l \rangle)= \frac{2 \langle l \rangle-\sqrt{4 \langle l \rangle+1}+1}{2 \langle l \rangle} 
\end{equation}
it is easy to see that now $\epsilon\in[0,1]$ for each value of $\langle l \rangle$, and that one can associate the parameter $\epsilon$ to how far back one wants to weight this parameter. 

One might argue that the entropy at $\epsilon=1$ diverges, but if one plots the entropy as a function of $\langle l \rangle$ instead.

\begin{equation}
\ ^{*} \vec S_{\langle l \rangle} =\frac{ 1}{I-  (\frac{2 \langle l \rangle-\sqrt{4 \langle l \rangle+1}+1}{2 \langle l \rangle}
)M} \ ^1\vec S 
\label{resolventl}
\end{equation}

One can easily show that for $\lim_{\langle l\rangle \rightarrow \infty} \ ^{*} \vec S_{\langle l \rangle}$ due to the fact that the number of paths grows as $r^{\langle l\rangle}$ for a constant $r$ in the transient phase, but when thermalization occurs, the entropy grows linearly with $\langle l \rangle$. For large values of $\langle l \rangle$, this entropy grows at $1/\sqrt{\langle l \rangle}$, given by:
\begin{equation}
\lim_{\langle l \rangle \rightarrow \infty}\ ^{*} \vec S_{\langle l \rangle} \approx \frac{ 1}{I-  (1-\frac{1}{ \sqrt{\langle l \rangle}}
)M} \ ^1\vec S + O(\frac{1}{\sqrt{\langle l \rangle}})
\label{resolventll}
\end{equation}
This indeed shows that a the complexity of a node $i$, for $\langle l \rangle \gg 1$, diverges as $a_i \sqrt{\langle l \rangle}$, with $a_i$ being a fitting parameter, which we hereon call \textit{asymptotic path complexity}. If one approximates the transient behavior of this entropy as a functio of the form $\ ^* S_{i}\approx a^1_i \sqrt{\langle l \rangle}+ a^0_i$, it is clear that the transient is characterized by the ratio $\frac{a^1_i}{a^0_i}$.

Before studying the growth of this entropy, we first write few identities. We write $ \ ^{*} \vec S_{\epsilon_1}=\frac{1}{\epsilon_1}R(\frac{1}{\epsilon_1}, M) \ ^{1}\vec S$, with $R(a,M)=\frac{1}{a I-M}$ then, one can use the operatox $x R(x,M)$ and use the second resolvent identity $R(x,M) R(y,M)=-(x-y)[R(x,M) -R(y,M)]$, to put in relation the two:

\begin{eqnarray}
\frac{1}{\epsilon_2} R(\frac{1}{\epsilon_2}, M)\ ^{*} \vec S_{\epsilon_1}=\frac{1}{\epsilon_2} R(\frac{1}{\epsilon_2}, M)  \frac{1}{\epsilon_1} R(\frac{1}{\epsilon_1}, M)\ ^1 \vec S  \nonumber \\
=-\frac{1}{\epsilon_2 \epsilon_1} (\frac{1}{\epsilon_2}-\frac{1}{\epsilon_1})[R(\frac{1}{\epsilon_2},M)-R(\frac{1}{\epsilon_1},M)]\ ^1 \vec S  \nonumber \\
=-\frac{1}{\epsilon_2 \epsilon_1} (\frac{1}{\epsilon_2}-\frac{1}{\epsilon_1})[\epsilon_2  \ ^* \vec S_{\epsilon_2}-\epsilon_1 \ ^* \vec S_{\epsilon_1}]
\end{eqnarray}
and thus find the identity:
\begin{eqnarray}
\epsilon_1 R(\frac{1}{\epsilon_2}, M)\ ^{*} \vec S_{\epsilon_1}=-(\frac{1}{\epsilon_2}-\frac{1}{\epsilon_1} ) [\epsilon_2\ ^{*}\vec S_{\epsilon_2}-\epsilon_1\ ^{*}\vec S_{\epsilon_1}]
\end{eqnarray}
which, after rearrangement can be casted into the simpler form:
\begin{equation}
\ ^* \vec S_{\epsilon_2}=\frac{\epsilon_1}{\epsilon_2}\frac{  \left(\epsilon _1 \epsilon_2 R(\frac{1}{\epsilon_2},M)+(\epsilon_2-\epsilon
   _1 )I\right)}{\left(\epsilon _2-\epsilon _1\right) } \ ^* \vec S_{\epsilon_1}
\end{equation}

showing that the $\epsilon$-Path complexity can be \textit{evolved} from one particular $\epsilon$ to another using the resolvent.

\subsection{Applications}
\begin{figure}
\centering
\includegraphics[scale=0.3]{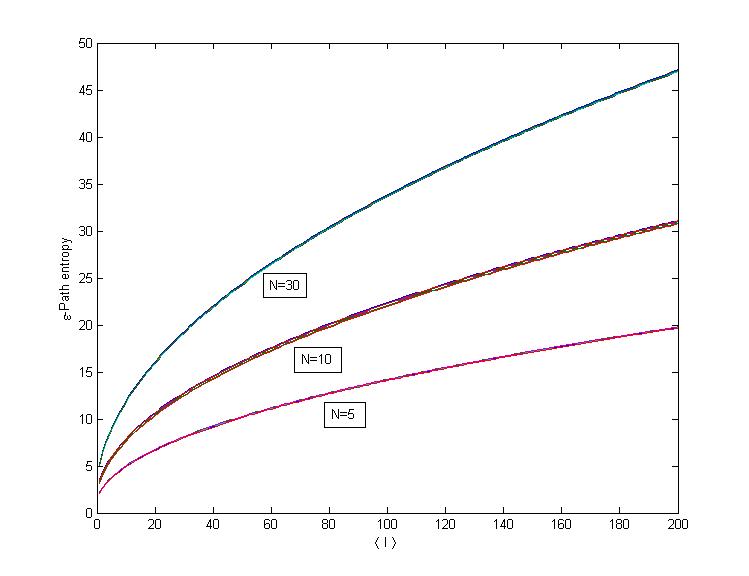} \includegraphics[scale=0.3]{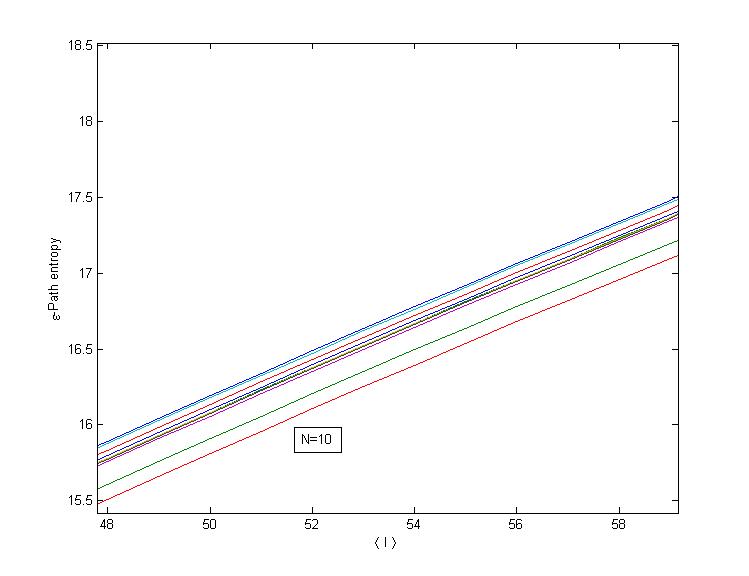}
\caption{\textit{Top:} Growth of entropies for various sizes of random weighted graphs. We observe that for long walks on graphs of different sizes, the entropy grows with different slopes. \textit{Bottom:} Although in left figure one can barely distinguish how the different node entropies grow, the right figure shows clearly that the entropies pertain the local microstructure.}
\label{fig:randomN}
\end{figure}

We are now interested in showing how the path entropy can indeed provide important informations of the properties of a graph. We thus study the random walk on a graph, with transition matrix $M=D^{-1} A$, with $D$ being the degree matrix. In particular, having shown that differently from analogous graph entropies present in the literature one can still distinguish asymptotically different nodes according to their path complexity, we would like to rank nodes according to their complexity for . A first test of this statement is applied to random (positive) matrices of different sizes $N$, as in Fig. \ref{fig:randomN}. As a first comment, it is easy to see that the complexity depends on the size of the graph, $N$, showing that different growth curves appear as a function of $\langle l \rangle$.  Although for different sizes, the entropies are clustered around similar values, zooming onto the curve shows that indeed these pertain the memory on their path complexity in the factor $a^1_i$, asymptotically. One can then perform a similar analysis for other type of random graphs.

\begin{figure}
\centering
\includegraphics[scale=0.4]{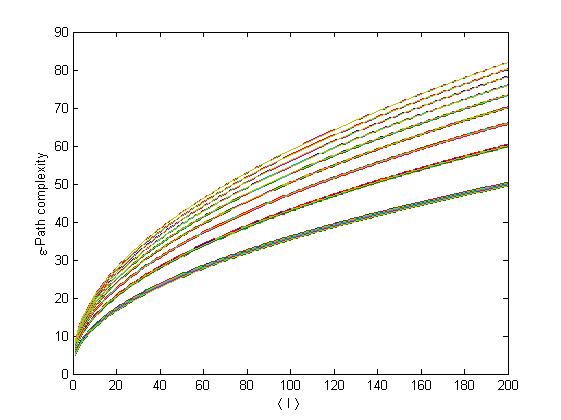}
\label{fig:random}
\caption{$\epsilon$-Path complexity for Erd\'{o}s-Renyi graphs, generated with value of the probability parameter $p=0.1,\cdots,0.9$. The lower set of curves is associated with the probability parameter $p=0.1$, and the higher with $p=0.9$.}
\label{erdosrenyi}
\end{figure}

We extend this analysis also to the case of  Erd\'{o}s-Renyi graphs in Fig. \ref{erdosrenyi}. We have generated various instances of random graphs, according to different realizations of the probability parameter $p$ to have a link or not; we have considered graph with the same number of nodes, $N=300$. It is easy to see tha different growth curves can be distinguished according to the parameter, although these curves become more and more similar for larger values of $p$.

\begin{figure}
\centering
\includegraphics[scale=0.34]{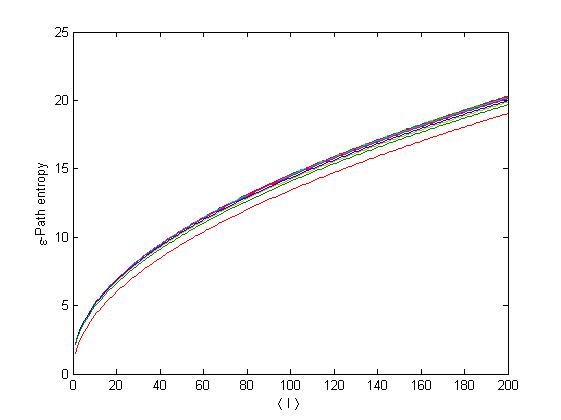} 
\includegraphics[scale=0.34]{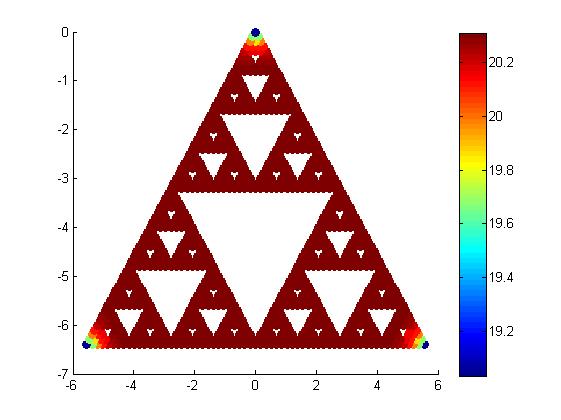}
\includegraphics[scale=0.34]{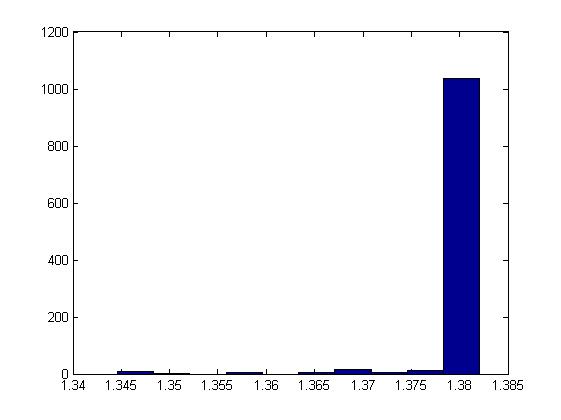}
\caption{\textit{Top:} Plot of the $\epsilon$-Path complexity for the nodes of a Sierpinsky fractal as a function of $\langle l \rangle$. We observe that, although most of the nodes have similar entropies, there are few outliers. In order to identify which nodes exhibit lower complexity, we plot a heatmap of a Sierpinsky fractal in the figure on the right. \textit{Central:} A heat map of path complexity for the Sierpinsky fractal evaluated at $\langle l \rangle=200$, and with a number of nodes $N=1096$. We observe how, although the self-similarity properties, the entropy is able to identify points which possess lower path complexity on the boundary. \textit{Bottom:} We plot the frequency distribution of asymptotic complexity. We see that a large fraction of node have analogous asymptotic path complexity value very close to 1.38, meanwhile few nodes, showin in the Top Right figure, take lower values.}
\label{fig:sier}
\end{figure}

As a case study, we evaluate the complexity of nodes on a self-similar graph, as for instance the Sierpinsky fractal. The results are shown in Fig. \ref{fig:sier}. We have analyzed the growth curves for each node, for a graph with $N=1095$ nodes, observing that few nodes exhibited lower growth curved as compared to the others. By plotting a heat map of the node complexity on the fractal, one observes that the nodes at the boundary of the Sierpinsky fractal have lower path complexity. A histogram of the asymptotic complexity $a^1_i$ shows that most of the nodes exhibit a similar complexity, meanwhile fewer nodes can be clearly distinguished from the others.

\section{Conclusions}
In this paper we have introduced and studied the entropy associated with the number of paths originating at a node of a graph. Motivated by distinguishing the asymptotic behavior of non-local entropy defined on a graphs, and inspired by earlier studies on macroscopic entropies, we have obtained a closed formula for the path complexity of a node in a graph. This entropy can be thought as the centrality operator applied to the local definition of entropy of a node in a graph, and depends on an external constant, that we introduced in order to keep the entropy finite asymptotically. Although the entropy introduced in the present paper is a non-local, the extra parameter has a nice interpretation in terms of the average number of walks to be considered. This allows to study the average transient behavior of the entropy, and in particular to introduce the asymptotic path complexity, given by the constant which charachterizes the asymptotic behavior of the path complexity of a node. We have applied this entropy to studying (normalized) random matrices, random graphs and fractals, and in particular have shown that the overal complexity of a node depends in value on the size of the graph. For random graphs, we have shown that one that the average asymptotic behavior of a node depends on the value of the probability parameter $p$. In addition, we have shown that this entropy is able to distinguish points in the bulk of a fractal from those at some specific boundaries, showing that these have lower path complexity as compared to the others. In general, we have the feeling of having introduced a new measure of macroscopic complexity for graphs, based on the fact that the number of paths generating at a node can differ substantially depending where a node is located with respect to the whole graph. Given this non-local definition, one would expect that the path complexity can give important insights on the relevance of topological properties of networks in several of their applications.

In addition, we have compared this entropy to those introduced in the past, showing that this entropy contributes to the growing literature on graph entropies.
As a closing remark, we believe that this entropy has better asymptotic properties (long walks) as compared to those introduced so far, and thus can be used to study the properties of large graphs. 
\section*{Aknowledgments}
We would like to thank J. D. Farmer, J. McNerney and F. Caccioli for comments on an earlier drafting of this entropy. We aknowledge funding from ICIF, EPSRC: EP/K012347/1.\\\ \\




\end{document}